\documentclass[10pt,letterpaper,twocolumn]{article} 
\usepackage{ol2}
\usepackage[draft]{hyperref}
\usepackage{amsmath}

\begin{document}

\twocolumn[ 
\title{Light self-trapping in a large cloud of cold atoms}
\author{Guillaume Labeyrie$^{*}$ and Umberto Bortolozzo }
\address{
Universit\'e de Nice-Sophia Antipolis, Institut Non Lin\'eaire de Nice, CNRS,\\ 
1361 route des Lucioles, F-06560 Valbonne, France
\\
$^*$Corresponding author: guillaume.labeyrie@inln.cnrs.fr
}

\begin{abstract}
We show that, for a near-resonant propagating beam, a large cloud of cold $^{87}$Rb atoms acts as a saturable Kerr medium and 
produces self-trapping of light. By side fluorescence imaging we monitor the transverse size 
of the beam and, depending on the sign of the laser detuning with respect to the atomic transition, we observe 
self-focusing or -defocusing, with the waist remaining stationary for an appropriate choice of parameters. 
We analyze our observations by using numerical simulations based on a simple 2-level atom model.
\end{abstract}


 ] 

\noindent In 1974, Bjorkholm and Ashkin observed the first optical spatial soliton in a bulk nonlinear 
medium, namely a hot sodium vapor~\cite{Bjorkholm1974}. Indeed, it was previously argued~\cite{Chiao1964} that 
the diffraction associated to a finite transverse size could be balanced by self-focusing e.g. in a Kerr 
medium, where the intensity profile of the laser beam induces a positive lens. This \textit{self-trapped} beam can 
propagate in a quite stable fashion if the nonlinearity is saturable. Spatial solitons have, since then, been studied in 
a variety of nonlinear media~\cite{Stegeman1999}, including, for instance, glass~\cite{Aitchison}, photorefractive crystals~\cite{Segev}, liquid crystals~\cite{Assanto} and light-valves~\cite{Piccardi}. In the last decade, laser-cooled atomic vapors have become 
increasingly employed to study both linear and nonlinear light propagation. Compared to other media, cold atoms 
offer several advantages such as e.g. the possibility to manipulate the properties of the medium, the large nonlinearities 
at low light levels, and the simple direct modeling from the microscopic expression of the atomic 
susceptibility. A few preliminary studies~\cite{Labeyrie2003, Wang2004, Labeyrie2007} have addressed the issue 
of self-focusing and -defocusing in cold atomic samples, in the ``thin-lens'' regime where modifications of the 
beam profile \textit{inside} the atomic medium are negligible.

In this Letter, we present the first experimental evidence of light self-trapping inside a cloud of laser-cooled atoms. 
This is achieved by sending a near resonant beam to propagate inside a large cloud of cold $^{87}$Rb atoms. The solitonic behavior is detected by monitoring the size of the beam through fluorescence side-imaging. After a brief description of the setup, we present our observations on the self-focusing and -defocusing regime and we show that there exists a regime of parameters for which the beam size remains stationary. The results are supported and discussed with the help of numerical simulations.
 
\begin{figure}[h!]
\centerline{\includegraphics[width=\columnwidth]{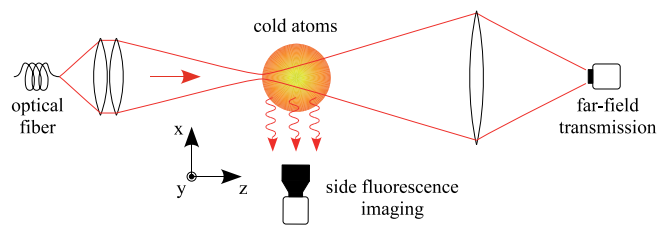}}
\caption{(Color online) Experimental setup. We focus a laser beam (waist $20~\mu$m) at the front side of the cold atomic cloud, and monitor its transverse size as it propagates using side fluorescence imaging. The far-field distribution of the transmitted beam is also recorded.}
\label{fig1}
\end{figure}

The experimental setup is sketched in Fig.~\ref{fig1}. Observing self-trapping requires the length of the 
nonlinear medium to be quite larger than the Rayleigh length of the incoming laser beam. 
Moreover, the spatial atomic density should be large enough to yield a sizable index of refraction. 
Fulfilling both criteria with cold atoms represents quite a challenge, requiring trapping and cooling a 
large number of atoms. We prepare our sample of cold atoms in a ``giant'' vapor-loaded MOT using six 
trapping beams of waist 3 cm with a total power of 600 mW, and containing roughly $10^{11}$ atoms at a 
temperature of $\approx 100 \mu$K. Because of light-induced inter-atomic repulsion, the cloud size is 
large with a quasi-Gaussian density profile of peak value $\rho_0 = 8 \times 10^{10}$ cm$^{-3}$. Along 
the beam propagation axis $z$, the FWHM size of the cloud is 11 mm, much larger than the probe's Rayleigh 
length $z_R = 1.6$ mm (beam waist $w_{in} = 20~\mu$m at the front entrance of the cloud). Once the MOT is 
loaded (within 1 s), we switch off the trapping lasers and magnetic field and shine the probe beam at the atoms 
for a $2~\mu$s duration. This short time is chosen to avoid nonlinear effects linked to optical pumping or 
mechanical effects~\cite{Labeyrie2007}. We checked that increasing the pulse duration up to $10~\mu$s does 
not affect our observations. In these conditions an atom scatters at most 38 photons during the 
interaction with the laser beam. We record from the side the fluorescence profile induced by 
the beam, using an EMCCD (Andor iXon+). The image is accumulated for up to 100 
cycles. The far-field distribution of the transmitted beam is recorded using another CCD camera.

\begin{figure}[h!]
\centerline{\includegraphics[width=0.87 \columnwidth]{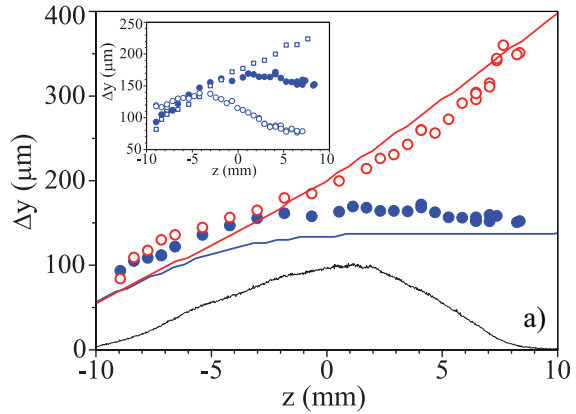}}
\caption{(Color online) Observation of self-trapping. We plot the fluorescence transverse size $\Delta y$ versus position $z$ in the atomic cloud. We compare symmetric detunings: $\delta = 5 \Gamma$ (solid circles) and $\delta = -5 \Gamma$ (open circles). The beam waist is $20~\mu$m, positioned at $z \approx -11.5$ mm, and the laser power is $100~\mu$W. The bold lines are numerical simulations (see text). The curve at bottom is the measured atomic density profile. The inset compares the measured data for $\delta = 3 \Gamma$ (open circles), $\delta = 5 \Gamma$ (filled circles) and $\delta = 7 \Gamma$ (open squares).}
\label{fig2}
\end{figure}

Fig.~\ref{fig2} shows an evidence of self-trapping of the laser beam inside the atomic cloud. We plot the transverse FWHM size $\Delta y$ (as obtained from a Gaussian fit) of the fluorescence induced by the probe, as a function of $z$. We compare the propagation for opposite detunings from the atomic resonance $\delta = \omega_l - \omega_{at}$, where $\omega_l$ and $\omega_{at}$ are the laser and atomic angular frequencies, respectively. The solid and open circles correspond to $\delta = 5\Gamma$  and $-5\Gamma$ respectively, where $\Gamma = 2 \pi \times 6.06$ MHz is the natural linewidth. The laser power is $100~\mu$W. The measured transmissions are $26 \%$ for $\delta > 0$ and $18 \%$ for $\delta < 0$. The bold curves are obtained from numerical simulations (see below) whereas the curve at bottom is the measured atomic density profile. The observed behavior clearly departs from linear propagation and shows a strong dependence on the sign of the detuning. For $\delta > 0$, after an initial growth the beam size remains constant, close to the expected behavior for a self-trapped beam when self-focusing exactly cancels diffraction. This behavior is emphazised in the inset, where we compare situations where self-focusing overcomes diffraction ($\delta = 3 \Gamma$, empty circles), balances it ($\delta = 5 \Gamma$, solid circles) or is dominated by it ($\delta = 7 \Gamma$, empty squares).

From Fig.~\ref{fig2} we see that self-focusing takes place on typical distances of the order of the cloud size, hence the thin-lens description~\cite{Labeyrie2003, Wang2004, Labeyrie2007} cannot be used here. A quantitative analysis requires full numerical simulations of the nonlinear propagation inside the cloud. It relies on a 2-level model, valid in the absence of optical pumping (Zeeman or hyperfine) and mechanical effects~\cite{Labeyrie2007}, which is justified for our short pulse duration. In the absence of Doppler broadening, the refractive index $n$ and photon scattering rate per atom $\Gamma^{\prime}$ are given by 
\begin{gather}
\label{index}
n(I, \delta) = 1 - g \frac{3 \lambda^3}{4 \pi^2}\rho \frac{\delta/\Gamma}{1+4(\delta/\Gamma)^2 + g I/I_{sat}}\\
\label{scatter}
\Gamma^{\prime}(I, \delta) = \frac{\Gamma}{2} \frac{g I/I_{sat}}{1+4(\delta/\Gamma)^2 + g I/I_{sat}} = \frac{\Gamma}{2} \frac{s}{1 + s} 
\end{gather}
where $s(I, \delta) = g (I/I_{sat}) / [1+4(\delta/\Gamma)^2]$ is the saturation. Here, $g = 7/15$ is the degeneracy factor accounting for the presence of several equally-populated Zeeman sub-levels, $\rho$ the atomic density, $I$ the laser intensity and $I_{sat} = 16$ W/m$^2$ the saturation intensity. The dispersive behavior of the refractive index (Eq.~\ref{index}) yields self-focusing and -defocusing for $\delta > 0$ and $< 0$, respectively. The photon scattering rate (Eq.~\ref{scatter}) is responsible for the attenuation of the laser beam as it propagates through the sample. The measured fluorescence signal is $S(y, z) \propto \rho(z) \int \Gamma^{\prime}(x, y, z)dx$, where we assume that the transverse atomic density profile is uniform on the scale of the beam size. Thus, the fluorescence signal does not directly reflect the ``beam size'' $I(y, z)$, except in the linear case $s \ll 1$. In addition, the scattered photons generated in the laser beam have to propagate through the atomic cloud to reach the CCD. This propagation is linear outside the beam, but \textit{diffusive} if the optical thickness of the cloud along $x$ is $>1$. Assuming scattered photons at the laser frequency, this condition is fulfilled for $\left | \delta \right | > 3 \Gamma$. Indeed, we typically observe a diffusive pedestal on our fluorescence beam profiles for $\left| \delta \right | < 4 \Gamma$. 

A qualitative picture for self-trapping can be provided assuming a ``step-like'' transverse profile for the refractive index. This assumption is reasonable for a uniform circular illumination~\cite{Chiao1964}, or for a Gaussian beam at high saturation $s \gg 1$~\cite{Labeyrie2007}. If the refractive index at beam center is larger than that in the wings, which is the case for $\delta > 0$ (Eq.\ref{index}), a critical angle $\theta_c$ for total internal reflexion can be defined. Thus, self-trapping can occur if the incident beam divergence $\theta_d = \lambda/\pi w_{in}$ satisfies
\begin{equation}
\label{guiding}
\theta_d < \frac{\pi}{2}- \theta_c \approx \sqrt{g \frac{3\lambda^3}{2\pi^2} \rho \frac{\delta/\Gamma}{1+4(\delta/\Gamma)^2}}.
\end{equation}
This inequality is independent of $I$ since we assumed $s \gg 1$. For $w_{in} = 20~\mu$m and $\delta = 5 \Gamma$, self-trapping occurs for $\rho > 8 \times 10^{10}$ cm$^{-3}$. In addition to assuming a step-like profile ($s \gg 1$), this simple approach neglects two important features of our system. First, the density profile (and thus the index step) is not uniform along the propagation direction. This implies that the condition above can be satisfied only in a limited spatial range around the center of the cloud, which explains the initial increase of the beam size for $\delta > 0$ in Fig.~\ref{fig2}. Second, with our parameters the attenuation of the beam is rather large due to scattering. This attenuation limits the range of $z$ where the condition $s \gg 1$ can be sustained.  

A numerical simulation is performed, including the non-uniform density profile of the medium along $z$ and the beam 
attenuation due to photon scattering. We employed a split-step method, computed the intensity distribution 
$I(x, y, z)$ and, then, the fluorescence signal and the far-field distribution of the output beam. 
The bold curves in Fig.~\ref{fig2} show the behaviors obtained with the measured experimental parameters, except 
for the peak atomic density ($1.1 \times 10^{11}$ cm$^{-3}$ here) which was increased by a factor $1.3$ to compensate 
for a systematic discrepancy between experiments and simulations. Such an uncertainty of $30 \%$ on the peak density is reasonable, since this quantity is difficult to determine accurately for large MOTs because of high optical thicknesses and non-Gaussian profiles. The agreement is then quite good both qualitatively and quantitatively.

\begin{figure}[h!]
\centerline{\includegraphics[width=0.9 \columnwidth]{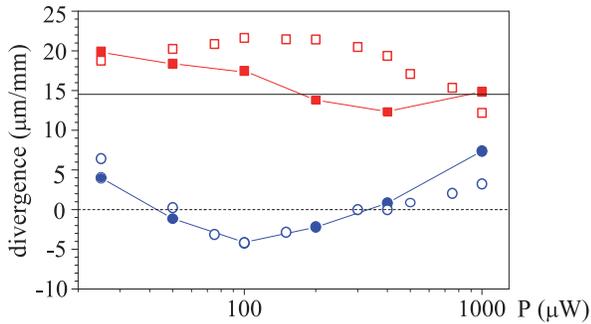}}
\caption{(Color online) Nonlinear propagation versus laser power. The divergence (see text) is plotted vs the laser power, 
for $\delta = 4 \Gamma$ (circles) and $\delta = - 4 \Gamma$ (squares). 
The filled and empty symbols correspond to experiments and simulations respectively.}
\label{fig3}
\end{figure}

\begin{figure}
\centerline{\includegraphics[width=0.8\columnwidth]{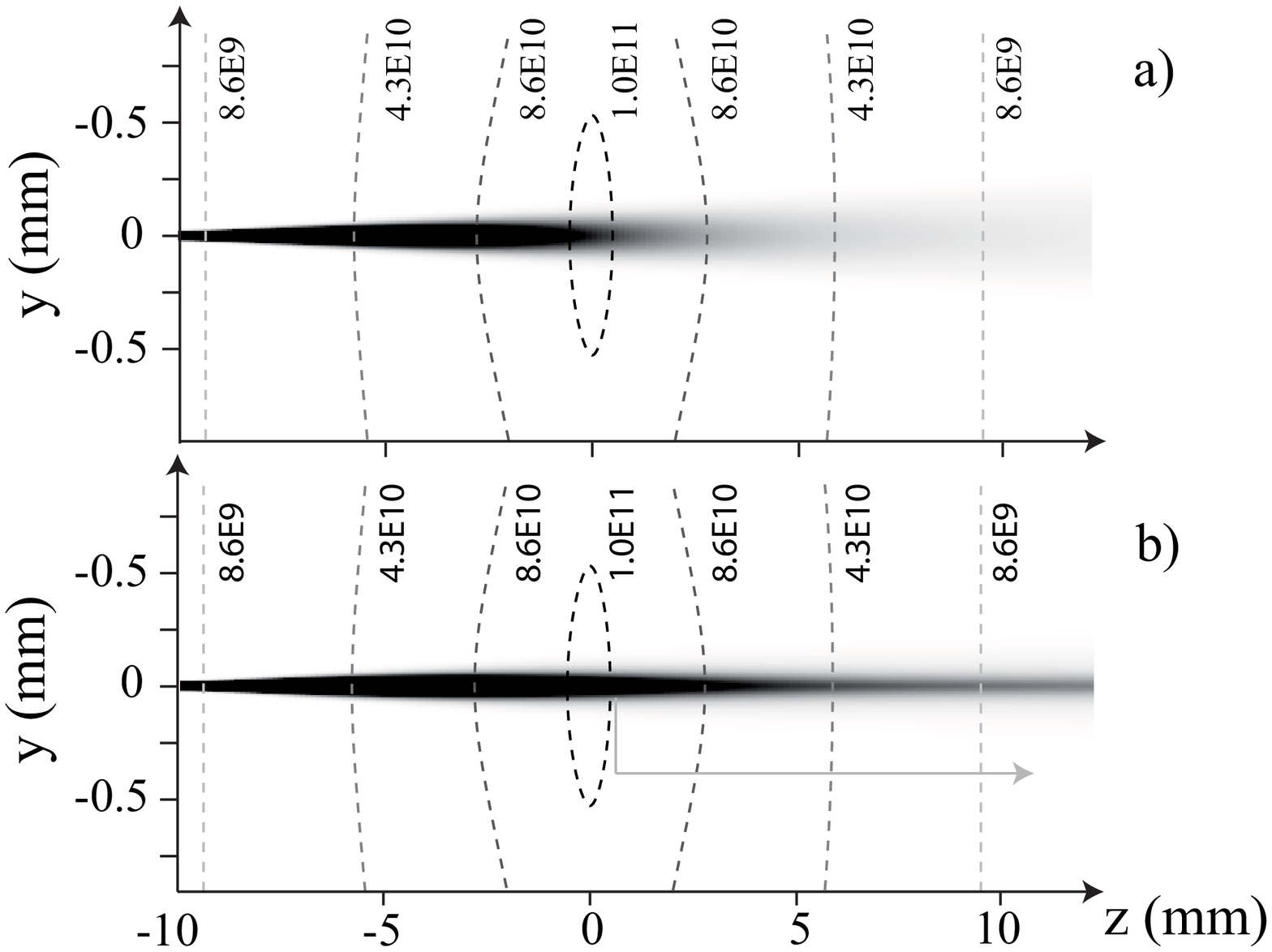}}
\caption{Numerical plots of the light intensity in the $y-z$ plane for the parameters 
of Fig.~\ref{fig2}: a) self-defocusing and b) self-focusing case; the contour lines represent the 
density of the cold atom cloud in cm$^{-3}$. The gray arrow indicates the region where the beam waist is constant.}
\label{fig4}
\end{figure}

We investigated the nonlinear propagation for $2 \leq \left| \delta \right| / \Gamma \leq 12$ and $25~\mu$W $\leq P \leq 1$ mW. A general feature is that the nonlinear effects, heralded by 
diverging behaviors for opposite $\delta$s, start to develop roughly $2$ mm before the peak density is 
reached (see Fig.~\ref{fig2}). The subsequent behavior for $z > 0$ can then be characterized by 
the ``divergence'' $\alpha = d(\Delta y) / dz$. In Fig.~\ref{fig3} we report $\alpha$ \textit{vs} the laser 
power, for $\left| \delta \right| / \Gamma = 4$. Circles and squares correspond to $\delta > 0$ and $< 0$, respectively (filled and empty symbols for experiment, respectively, simulation). The horizontal solid line represents the divergence of the incident Gaussian beam. The dotted line $\alpha = 0$ corresponds to self-trapping. For $\alpha < 0$ the beam is strongly focused inside the cloud, while for $\alpha > 0$ the beam diverges, either due to insufficient self-focusing ($\delta > 0$) or self-defocusing ($\delta < 0$). The comparison between experiment and numerics is quite satisfying.
In Fig.~\ref{fig4}a) and b) we show numerical plots of the propagating light intensity in the $y-z$ plane 
for the parameters of Fig.~\ref{fig2}. The contour lines represent the density of the cold atom cloud. 
In the self-focusing case b), a stable self-trapping of the beam is observed in the region shown by the gray arrow.

In conclusion, we have provided an experimental evidence of light self-trapping in a large cloud of cold atoms. Numerical simulations based on the microscopic nonlinear atomic susceptibility are in good agreement with the experiment. 
The observation of such a solitonic behavior paves the way for the study of nonlinear physics with light in cold 
atoms, a medium with unique possibilities in terms of tunability. In particular, this work constitutes a first 
step towards the study of optical localized structures in cold atoms. 
In contrast with the situation in hot vapors~\cite{Schappers2000}, one expects here a strong interaction between 
the localized structures and the atomic external degrees of freedom, with a consequent complex dynamics~\cite{Saffman1998}.

G.L. thanks T. Ackemann and R. Kaiser for stimulating discussions. This work was supported by CNRS, Universit\'e de Nice-Sophia Antipolis and R\'egion PACA.

\pagebreak

\section*{Informational Fourth Page}
In this section, please provide full versions of citations to
assist reviewers and editors (OL publishes a short form of
citations) or any other information that would aid the peer-review
process.


\begin{thebibliography}{99}

\bibitem{Bjorkholm1974} J.E. Bjorkholm and A. Ashkin, "cw Self-Focusing and Self-Trapping of Light in Sodium Vapor", Phys. Rev. Lett. \textbf{32}, 129-132 (1974).
\bibitem{Chiao1964} R.Y. Chiao, E. Garmire and C.H. Townes, "Self-Trapping of Optical Beams", Phys. Rev. Lett. \textbf{13}, 479-482 (1964).
\bibitem{Stegeman1999} G. I. Stegeman and M. Segev, "Optical Spatial Solitons and Their Interactions: Universality and Diversity", Science \textbf{286} 1518-1523 (1999).
\bibitem{Aitchison} J. S. Aitchison, A. M. Weiner, Y. Silberberg, M. K. Oliver, J. L. Jackel,
D. E. Leaird, E. M. Vogel and P. W. Smith, "Observation of spatial optical solitons in a nonlinear glass waveguide", Opt. Lett. {\bf 15}, 471-473 (1990).
\bibitem{Segev} M. Segev, B. Crosignani, A. Yariv and B. Fischer, "Spatial solitons in photorefractive media", Phys. Rev. Lett. {\bf 68}, 923-926   (1992).
\bibitem{Assanto} M. Peccianti, C. Conti, G. Assanto, A. De Luca, and C. Umeton, "Routing of anisotropic spatial solitons and modulational instability in liquid crystals", Nature {\bf 432}, 733-737 (2004).
\bibitem{Piccardi} A. Piccardi, U. Bortolozzo, S. Residori and G. Assanto, "Spatial solitons in liquid-crystal light valves", Opt. Lett. {\bf 34}, 737 -739 (2009).
\bibitem{Labeyrie2003} G. Labeyrie, T. Ackemann, B. Klappauf, M. Pesch, G.L. Lippi and R. Kaiser, "Nonlinear beam shaping by a cloud of cold Rb atoms", Eur. Phys. J. D \textbf{22}, 473-483  (2003). 
\bibitem{Wang2004} Y. Wang and M. Saffman, "Experimental study of nonlinear focusing in a magneto-optical trap using a Z-scan technique", Phys. Rev. A \textbf{70}, 013801-9 (2004).
\bibitem{Labeyrie2007} G. Labeyrie, G.L. Gattobigio, T. Chaneli\`ere, G.L. Lippi, T. Ackemann, and R. Kaiser, "Nonlinear lensing mechanisms in a cloud of cold atoms", Eur. Phys. J. D \textbf{41}, 337-348 (2007).
\bibitem{Schappers2000} B. Sch\"apers, M. Feldmann, T. Ackemann, and W. Lange, "Interaction of Localized Structures in an Optical Pattern-Forming System", Phys. Rev. Lett. \textbf{85}, 748-751 (2000).
\bibitem{Saffman1998} M. Saffman, "Self-Induced Dipole Force and Filamentation Instability of a Matter Wave", Phys. Rev. Lett. \textbf{81}, 65-68 (1998).


\end{thebibliography}
\end{document}